\documentclass{article}
\usepackage[T1]{fontenc} 
\usepackage[utf8]{inputenc} 
\usepackage{ismir,amsmath,cite,url}

\usepackage{bbm}

\usepackage{graphicx}
\usepackage{color}
\usepackage{tabularx}
\usepackage{adjustbox}
\usepackage{lipsum}
\usepackage{multirow}
\usepackage{cellspace}
\usepackage[utf8]{inputenc}
\usepackage{algorithm}
\usepackage{algorithmic}
\usepackage{arevmath}     
\usepackage{subcaption}
\usepackage{mwe}

\usepackage{pict2e}
\usepackage{xcolor} 

\makeatletter
\DeclareRobustCommand{\onontimes}{%
  \mathbin{\mathpalette\on@ntimes\relax}%
}
\newcommand{\on@ntimes}[2]{%
  \vcenter{\hbox{%
    \sbox0{\m@th$#1\otimes$}%
    \setlength\unitlength{\wd0}%
    \begin{picture}(1,1)
    \linethickness{0.35pt}
    \put(.5,.5){\circle{.8}}
    \end{picture}%
  }}%
}

\newcommand\Tstrut{\rule{0pt}{2.9ex}}         
\newcommand\Bstrut{\rule[-1.2ex]{0pt}{0pt}}   
\newcommand\TBstrut{\Tstrut\Bstrut}           
\usepackage{lineno}

\title{Open Set Recognition for Music Genre Classification}
 
\threeauthors
   {Zhanlin Liu} {Warner Music Group \\ { Kevin.Liu@wmg.com}}
   {Julien De Mori} {Warner Music Group \\julien.demori@wmg.com}
   {Kobi Abayomi} {Seton Hall University \\ kobi.abayomi@shu.edu}

\def\authorname{Z. Liu, J. De Mori, and K. Abayomi}

\begin{document}
\maketitle
\begin{abstract}
Music genres, prototypically, are poorly defined. Recent work suggests a path - via open set recognition - to identify unknown genres solely on features extracted from audio. We explore segmentations of known and unknown genre classes using the open source GTZAN and FMA datasets; for each, we begin with `best-case' closed set genre classification, then we apply open set recognition methods. This allows us, on audio samples from unknown genres, to establish a baseline capacity for novel genre detection. This design, as well, c
an illustrate interaction between genre labeling  and open set recognition accuracy.
\end{abstract}

\section{Introduction}\label{sec:introduction}

Aural boundaries between music genres, \textit{e.g. labeled sets} -- Classical , Jazz, Hip-hop, Rock, alternative, Experimental, etc -- are imprecise and poorly defined. The antecedents for the ill defined labels differ: Classical music is characterized by a Europeanorigin \cite{jeremy}; so-called Pop is a function of a time-indexed neighborhood of popularity - local audience affinities. Experimental music, in contrast, can be defined \textit{against} this context. This paper explores these boundaries and their time-dependent definitions via a comprehension of the labeled sets of genre as functions of measurable aural features implicitly capturing instrumentation, harmony, melody, chord progressions, vocal qualities, etc.

The GTZAN \cite{tzanetakis2002musical} and Free Music Archive (FMA) \cite{defferrard2016fma} datasets have been used to benchmark the genre identification task via aural features. Genre classification performance on this data also illustrates the potential utility deep learning models provide on this classification problem \cite{ghildiyal2020music, kong2020panns, liu2021bottom, chillara2019music}, commonly via supervised learning on known, labeled sounds. 

A traditional supervised genre model, thus, can only classify music as of a known type - i.e. as a genre already established and observable in the labeled training data. This precludes classification as a novel type, e.g. predicting a particular sound or sound collection comprising an unknown class. A classifier's ability to identify sounds in exclusion from \textit{every} class suggests its potential to detect a new, novel genre. Colloquially, sounds from this novel class should be unrecognizable from any extant labeled class.




Open set recognition (OSR) has been demonstrated to be useful for detecting unknown classification categories in various domains of application, including  image classification \cite{scheirer2011meta, neal2018open, kong2021opengan}, acoustic event classification \cite{naranjo2020open, battaglino2016open}, and text classification \cite{venkataram2018open, shu2017doc}. Machine learning algorithms such as Support Vector Machines \cite{battaglino2016open, scheirer2014probability, jain2014multi}, Nearest Neighbour classifiers \cite{mendes2017nearest} and Sparse Representation \cite{zhang2016sparse} can be modified to address this OSR problem. Performance of these traditional machine learning models is unsatisfactory in big data settings: more specifically when the number of observations in each class is large. Recent advances in deep learning based solutions for OSR using SoftMax with threshold and OpenMax \cite{bendale2016towards}, though, suggest a more scalable approach to novel genre identification.

This paper offers a method for the novel genre identification problem that both
\begin{itemize}
    \item classifies audio into existing music genres (those available in the labeled training data), and
    \item  classifies audio into new music genres (those \textit{not} present in the labeled training data).
\end{itemize}

We also conduct an experimental analysis and exploration of the various factors of this methodology that impact open set genre classification accuracy.


The remainder of the paper is organized as follows. Section ~\ref{sec:method} introduces the OSR automatic system for music genre classification and presents how the closed set classifier and open set classifier are constructed. Section ~\ref{sec:experiment} illustrates the experimental setup as well as the results on both the GTZAN dataset and the FMA dataset. Section ~\ref{sec:con} concludes the paper with a discussion on the results and future research directions. 
\section{Methodology}\label{sec:typeset_text}
\label{sec:method}
\begin{figure*}[!ht]
{\centering
\includegraphics[width=0.98\textwidth]{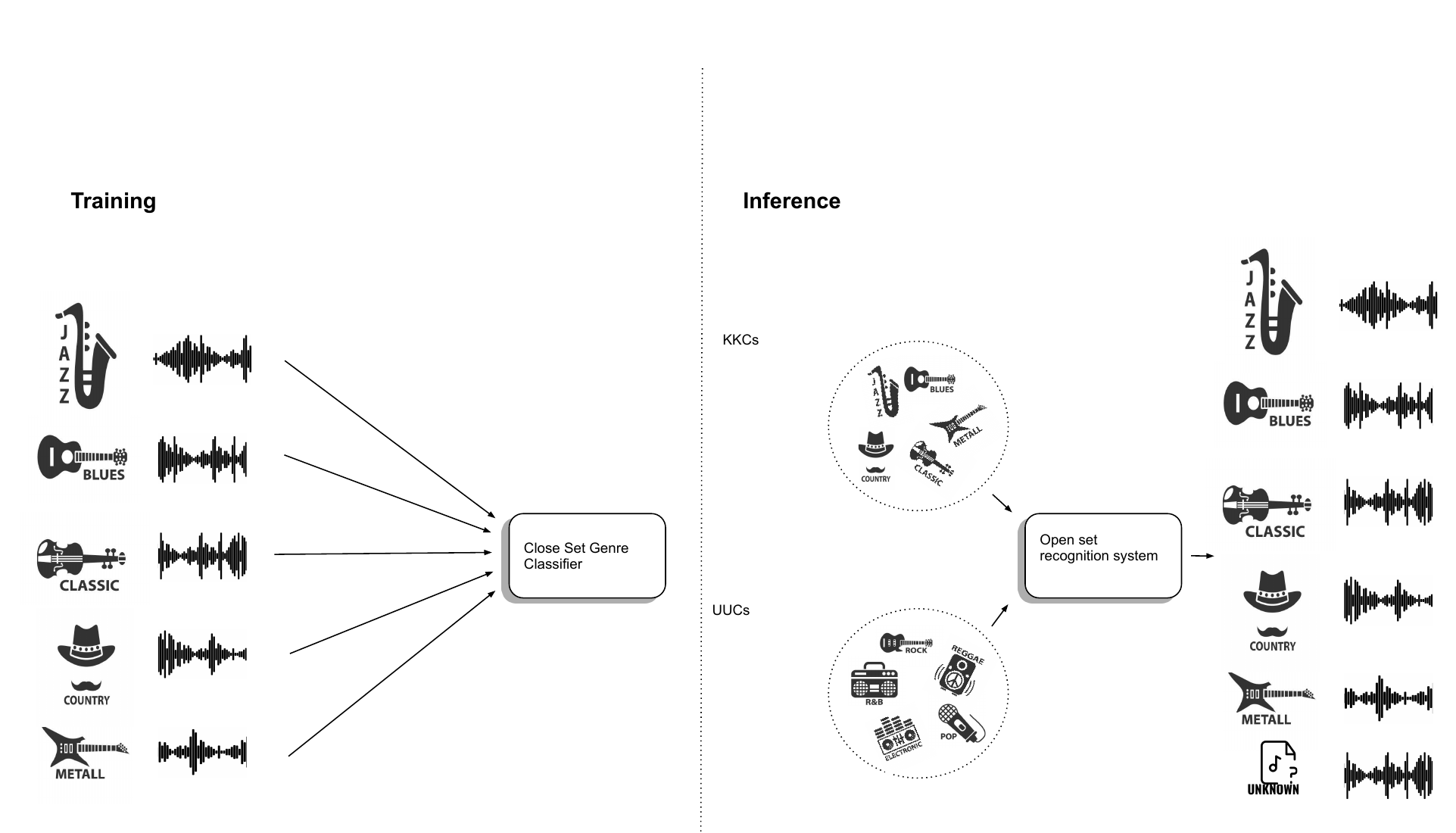} \\
\centering\caption{\small{Training and inference processes for the OSR pipeline for genre classification. In the training stage (left), a closed set classifier is trained on the KKCs. In the inference stage (right), an OSR system classifies samples into the KKCs and an additional class for cases in which the sample is deemed unlike any of the KKCs}}
\label{fig:pipeline}
}
\end{figure*}
Based on the definition in \cite{geng2020recent}, in a classification task there are \emph{known known classes} (KKCs), \emph{known unknown classes} (KUCs), \emph{unknown known classes} (UKCs), and \emph{unknown unknown classes} (UUCs). In this paper, we formulate the problem of open set recognition for music genre classification with only KKCs and UUCs, where the KKCs represent the closed set genre labels that are observed in training the deep classification model and UUCs represent the open set genre labels that do not appear in the training process. Colloquially, the closed set model knows nothing of the open set data (UUCs) that it runs inference on after training. Figure \ref{fig:pipeline} shows the framework of the open set recognition pipeline including training and inference steps. As illustrated, a closed set genre classifier is trained only on samples from the KKCs. An OSR algorithm is trained and utilized in the inference process to classify music audio as one of the KKCs or an unknown class. In this section, we will illustrate the closed set genre classifier and the open set recognition model used in this paper.  



\subsection{Closed Set Genre Classifier}\label{subsec:body}
Kong et al. proposed a set of architectures
in 2020 \cite{kong2020panns} that utilizee transfer learning and fine-tuning of model weights trained on the AudioSet dataset \cite{gemmeke2017audio}. CNN14, which is one of the proposed model architectures, has achieved state-of-the-art performance in multiple audio classification tasks including genre classification on the GTZAN dataset. Therefore, we propose using a CNN14 architecture and transfer learning as the closed set genre classifier. The model architecture using transfer learning based on the CNN14 architecture is illustrated in Table \ref{tab:cnn14}. 

\begin{table}[!ht]
\renewcommand{\arraystretch}{1.0}
  \centering
  \begin{adjustbox}{width=0.3\textwidth}
  \begin{tabular}{c}
  \hline
    Log-mel spectrogram  \\
\hline 
    $\big(3 \times 3 \; @ \; 64, BN, ReLU \big) \times 2 $ \\ 
\hline 
    Pooling $2 \times 2 $ \\ 
\hline 
$\big(3 \times 3 \; @ \; 128, BN, ReLU \big) \times 2 $ \\
\hline 
Pooling $2 \times 2 $ \\
\hline
$\big(3 \times 3 \; @ \; 256, BN, ReLU \big) \times 2 $ \\
\hline 
Pooling $2 \times 2 $ \\
\hline
$\big(3 \times 3 \; @ \; 512, BN, ReLU \big) \times 2 $ \\
\hline 
Pooling $2 \times 2 $ \\
\hline
$\big(3 \times 3 \; @ \; 1024, BN, ReLU \big) \times 2 $ \\
\hline 
Pooling $2 \times 2 $ \\
\hline
$\big(3 \times 3 \; @ \; 2048, BN, ReLU \big) \times 2 $ \\
\hline 
Global pooling \\ 
\hline
FC 2048 ReLU \\ 
\hline
FC $N$ ReLU \\ 
\hline
  \end{tabular}
  \end{adjustbox}
  \caption{Transfer learning in the CNN14 architecture. BN represents the batch normalization, FC represents a fully connected layer, and \emph{N} represents the number of KKCs. The model input takes a batch of 30-second audio waveforms.}
  \label{tab:cnn14}
\end{table}

\subsection{Open Set Classifier}
SoftMax with threshold and OpenMax are two methods commonly used to address the open set recognition problem. In this section, we describe the SoftMax with threshold method and OpenMax method in detail.

The softmax function is often used as the activation function in the output layer for DNNs in multi-class classification tasks. After applying the softmax function, each node of the output layer can be regarded as the probability of belonging to the corresponding class. It follows that a test sample is from an unknown class if the softmax function produces low probabilities for all known classes. By thresholding the output probabilities on the output layer, a test sample can be identified as an unknown class.

Alternatively, OpenMax, which was proposed in \cite{bendale2016towards} by Bendale et al, is regarded as pioneering work in deep open-set classifiers. It uses the extreme value theorem (EVT) \cite{smith1990extreme} and defines a per-class compact abating probability (CAP) model to reject unknown inputs with a threshold. 

To obtain OpenMax scores, activation vectors for each training instance $x$ are computed using the trained closed set classifier. After calculating the activation vectors for each training instance, $v_{1}(x), \ldots, v_{N}(x)$, \text{where} $N$ is the number of KKCs, can be obtained based on the activation vector for each training instance and its corresponding label. $\mu_{j}, j = 1, \ldots, N$ referred to as the mean activation vector (MAV), can be computed for each class separately by taking the mean over only correctly classified training examples. Afterwards, a Weibull distribution $\rho_{j} = (\tau_{j}, \lambda_{j}, \kappa_{j}),$ where $\tau_{j}, \lambda_{j}, \kappa_{j}$ are the parameters of the Weibull distribution, can be fitted to each class on the largest distances between the MAV class and positive training instances using the libMR \cite{scheirer2011meta} FitHight function. We use the euclidean distance as the measure of distance in this paper. After obtaining the activation vectors for each class, calculating the MAV, and fitting the Weibull distributions, the OpenMax probability can be estimated using the procedures in Algorithm \ref{OpenMax} \cite{bendale2016towards}. Three hyper-parameters are used in the OpenMax algorithm. Tail size ($\eta$) is used during the parameter estimation phase for fitting the Weibull distribution by holding out a small set of data samples. $\alpha$ is the parameter used in revising the probabilities for the "top" classes. $\epsilon$ is the threshold used to support the rejection of uncertain inputs. 

\begin{algorithm}[!ht]
\caption{OpenMax probability estimation with rejection of unknown or uncertain inputs.}
\begin{algorithmic}[1] \label{OpenMax} 
\REQUIRE Activation vector $\textbf{v(x)} = v_{1}(x), \ldots, v_{N}(x)$.
\\
\REQUIRE \textbf{means} $\mu_{j}$ and libMR models $\rho_{j} = (\tau_{i}, \lambda_{i}, \kappa_{i})$ \\
\REQUIRE $\alpha$, the number of "top" classes to revise \\
\STATE{Let $s(i) = \text{argsort}(v_{j}(x))$; Let $w_{j}=1$}
     \FOR{$i = 1, \ldots, \alpha $}
	\STATE $w_{s_{(i)}}(x) = 1 - \frac{\alpha -i}{\alpha}e^{-\big(\frac{\|x-\tau_{s(i)}\|}{\lambda_{s_{(i)}}}\big)^{\kappa_{s_{(i)}}}}$
\ENDFOR
\STATE{Revise activation vector $\hat{v}(x) = \textbf{v(x)}w(x)$ } \\
\STATE{Define $\hat{v}_{0}(x) = \sum_{i}v_{i}(x) (1-w_{i}(x))$} \\
\STATE{\[\hat{P}(y=j|x) = \frac{e^{\hat{v}_{j}(x)}}{\sum_{i=0}^{N}e^{\hat{v}_{i}(x)}}\]} \\
\STATE{Let $y^{\star} = \text{argmax}_{j}P(y==j|x)$} \\
\STATE{Reject input if $y^{\star} ==0$ or $P(y=y^{\star}|x) < \epsilon$}
\end{algorithmic}
\end{algorithm}

\section{Experimental Analysis}
\label{sec:experiment}
This section illustrates the experimental setup, including the datasets, the selection of classes in the closed and open space for each experiment and a discussion of the results.

The experimental design and setup are also aimed at providing insights into the impact that
\emph{data quantity},
\emph{closed and open set selection},
\emph{model selection},
\emph{data label quality}
have on the performance of the OSR music genre classifier.

\subsection{Datasets}\label{sec:page_size}
We evaluate the performance of the OSR algorithm for music genre classification on two benchmark datasets: the GTZAN dataset and the FMA dataset. Each dataset is split into a training set, an evaluation set, and a test set for each experiment. The training set and the evaluation set only contain samples from the KKCs while the test set also contains samples from genres in the UUCs. The KKCs and UUCs are defined in Table \ref{tab:performance}. More detail on the training and test split for each experiment is presented in Table \ref{tab:data_description}.

The GTZAN dataset consists of 1000 single-label audio files uniformly distributed among 10 genre labels. Each audio sample is a 30-second mp3. In this paper, we split the GTZAN dataset using the data split proposed in \cite{Charles2013}. The genre labels in the closed and open sets for Experiments 1-4 are illustrated in Table \ref{tab:performance}.
Notice from Table \ref{tab:data_description} that the training set and evaluation set has a smaller number of samples than the split proposed in \cite{Charles2013} due to the fact that the training and evaluation sets only contain genres in the closed set labels. 

For all experiments, the choice of KKCs and UUCs was made based on prior experiments and the evidence that improving the accuracy of the closed set classifier yields an improved open set classifier \cite{vaze2021open}. We chose genres for KKCs that are better defined by the properites of their audios, whose embeddings have relatively lower mean distance to the genre's centroid, whilst also having experiments with balanced amounts of data in the open and closed sets. 


The FMA dataset contains three different splits including FMA small, FMA medium, and FMA large. The descriptions can be found in \cite{defferrard2016fma}. 
We evaluate the model performance on the FMA dataset in three different experiments (index at 2, 3, and 4), where Table \ref{tab:data_description} describes the sample size for the training, the evaluation as well as the test set and Table \ref{tab:performance} illustrates the choices of KKCs and UUCs for each. We illustrate how to split data into the training set, the evaluation set, and the test set as follows:
\begin{enumerate}
    \item Experiment 2 uses 8 genres in the FMA small as KKCs to train and evaluate the closed set classifier and inference is run on UUCs from the remaining 8 genres in the valid split of the FMA medium dataset. 
    \item Experiment 3 redefines the KKCs and UUCs while still using the FMA small and medium datasets. 
    The evaluation and test sets use samples from both the FMA small and medium datasets. The samples of KKCs are split evenly for evaluation and testing.   
    \item Experiment 4 is the same as Experiment 3, with training and inference on audio samples from all FMA datasets including the FMA large split. Note that we only use audio sample with genre labels in the FMA large split. 
\end{enumerate}

\begin{table}[!ht]
\renewcommand{\arraystretch}{1.2}
  \centering
  \begin{adjustbox}{width=0.45\textwidth}
  \begin{tabular}{|c|c|c|c|}
  \hline
    Data & Training set & Evaluation set & Test set \\
\hline 
    GTZAN & 226 & 103  & 290  \\ 
\hline 

\multirow{2}{*}{FMA small \& medium} & 6400 & 800 & 2505 \\ 
   \cline{2-4}  
   &15040 &940 & 1565  \\
\hline
FMA large & 26045& 1516 & 3188\\
\hline
  \end{tabular}
  \end{adjustbox}
  \caption{This table contains the data sources as well as the number of samples for the training, evaluation, and test sets for each experimental setup. The size of the training and evaluation sets are based only on the audio samples in the KKCs defined in Table  \ref{tab:performance} respectively. }
  \label{tab:data_description}
\end{table}

\subsection{Objective Metrics}
The accuracy defined in \eqref{eqn:acc} is used to evaluate the model performance. Here $TP_{k}$ and $FN_{k}$ are the number of true positives and false negatives corresponding to class $k$. In the closed set model training process, $k \in [1, N]$, where $N$ is the number of KKCs. In the open set recognition process, $k \in [0, N]$ with UUCs indexed at $0$.  

\begin{equation}
    \frac{\sum_{k}TP_{k}}{\sum_{k}(TP_{k} + FN_{k})}.
    \label{eqn:acc}
\end{equation}

\subsection{Implementation Details}

As discussed in the previous section, we use transfer learning based on the CNN14 \cite{kong2020panns} architecture as the model backbone for the closed set classifier. In addition, the model weights are fine-tuned based on a task on the AudioSet dataset. For each experiment, we set the maximum epoch to 50. We compare the model performance by freezing or not freezing the CNN14 model weights. Moreover, we also compare the model performance with and without mixup augmentations for closed set classifier training. We find that the best model performance achieved for the GTZAN dataset occurs when the pre-trained model weights are frozen, and without mixup augmentation. However, better closed set model accuracy is achieved by not freezing the model weights and including mixup augmentation in the FMA experiments. All experiments use the Adam optimizer with a learning rate of 0.0001. 

For the OSR task, we use grid search for $\alpha \in [N/2, N]$ and $\eta$ ranges between 20 and 40 with increments of 5. The best model performances are presented in the next section.

\subsection{Results}
Table \ref{tab:performance} shows the best performance for each experimental setup with the optimally identified hyper-parameters including $\alpha$, $\eta$, and $\epsilon$. In the following section, we elaborate on the results summarized in the table. 



\begin{table*}[hbt!]
\renewcommand{\arraystretch}{1.5}
  \centering
  \begin{adjustbox}{width=0.99\textwidth}
  \begin{tabular}{|c|c|c|c|c|c|c|c|}
  \hline
    Index &Data & KKCs & UUCs & $Acc_{c}$ & $Acc_{s}$ & $Acc_{st}$ & $Acc_{ot}$\\
\hline 
    1 & GTZAN & Blues, Classical, Country, Jazz, Metal & Disco, Hip-hop, Pop, Reggae, Rock &0.829 & 0.417 &0.769 &0.724 \\ 
\hline 

2 & FMA small \& medium &  \shortstack{ \Tstrut Hip-hop, Pop, Folk, Experimental, Rock, \\
                    International,
                    Electronic, Instrumental} & \shortstack{Jazz, Soul-RnB, Blues,
                     Spoken, Country, \\ Classical, Old-Time Historic,
                     Easy Listening} &0.609 & 0.556 &0.556 &0.556 \\ 
   \hline
3 & FMA small \& medium   & \shortstack{ \Tstrut Hip-hop, Rock,
                    Electronic, Instrumental, Jazz, \\ Blues,
                    Spoken, Country, Classical, Old-Time  Historic} & \shortstack{ \Tstrut Pop, Folk, Experimental,\\
                    International,
                   Soul-RnB, Easy Listening}&0.717 &0.431&0.597 &0.608\\
\hline
4 & FMA large \TBstrut & \shortstack{ \Tstrut Hip-hop, Rock,
                    Electronic, Instrumental, Jazz, \\ Blues,
                    Spoken, Country, Classical, Old-Time  Historic} & \shortstack{ \Tstrut Pop, Folk, Experimental,\\
                    International,
                   Soul-RnB, Easy Listening}& 0.718 &0.343 &0.622 &0.635\\
\hline
  \end{tabular}
  \end{adjustbox}
  \caption{The table shows the model performance based on different datasets and different KKC and UUC selections. $Acc_{c}$ is the closed set test accuracy - accuracy on KKCs. $Acc_{s}$ is the overall accuracy using the softmax activation on the entire test set, including unknown classes. $Acc_{st}$ and $Acc_{ot}$ are the accuracies for the test set using SoftMax with thresholds and OpenMax with thresholds respectively. }
  \label{tab:performance}
\end{table*}

\subsection{Discussion}

We discuss the results and observe how they vary with the \emph{rejection threshold $\epsilon$}, \emph{open set model}, \emph{training data size}, \emph{open and closed set class selection}, \emph{data quality}, and observations on how well defined specific genres are based on the labels and audio features in these datasets.

\begin{figure}[hbt!]
\centering
\begin{subfigure}[hbt!]{0.45\columnwidth}
    \centering
    \includegraphics[width=\textwidth]{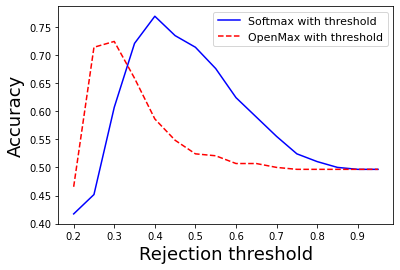}
    \caption{$\alpha=2, \eta = 20$.}
    \label{GTZAN_threshold}
\end{subfigure}
~ 
\begin{subfigure}[hbt!]{0.45\columnwidth}
    \centering
    \includegraphics[width=\textwidth]{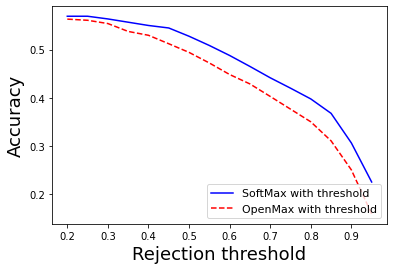}
    \caption{$\alpha=6, \eta = 35$.}
    \label{FMA_small}
\end{subfigure}
~ 
\begin{subfigure}[hbt!]{0.45\columnwidth}
    \centering
    \includegraphics[width=\textwidth]{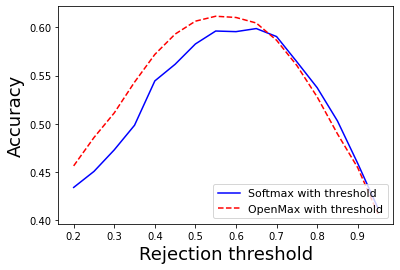}
    \caption{$\alpha=8, \eta = 30$.}
    \label{FMA_self}
\end{subfigure}
~ 
\begin{subfigure}[hbt!]{0.45\columnwidth}
    \centering
    \includegraphics[width=\textwidth]{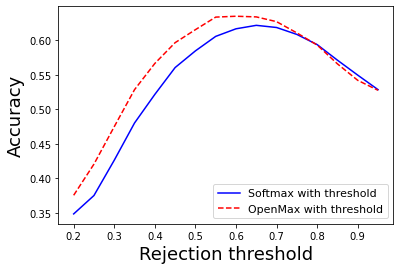}
    \caption{$\alpha=10, \eta = 35$.}
    \label{FMA_large}
\end{subfigure}
\caption{Accuracy vs. the rejection threshold for each experiment. (a) shows the result on Experiment 1. (b) and (c) present results using the FMA small \& medium dataset on Experiment 2 and Experiment 3, respectively (d) shows the result on Experiment 4.}
\label{fig:threshold}
\end{figure}

Figure \ref{fig:threshold} plots the accuracies of OpenMax and SoftMax for different rejection thresholds $\epsilon$. In plot (b), we notice that the highest accuracy occurs at the lowest $\epsilon$ amongst experiments. Note that for all experiments the function of accuracy appears concave with respect to the rejection threshold. The optimal accuracy reached in Experiment 2 is the lowest among all experiments, consistent with the finding in \cite{vaze2021open} that an closed set classifier performance is correlated to the accuracy on the open set recognition task.

\begin{figure}[hbt!]
\centering
\begin{subfigure}[hbt!]{0.48\columnwidth}
    \centering
    \includegraphics[width=\textwidth]{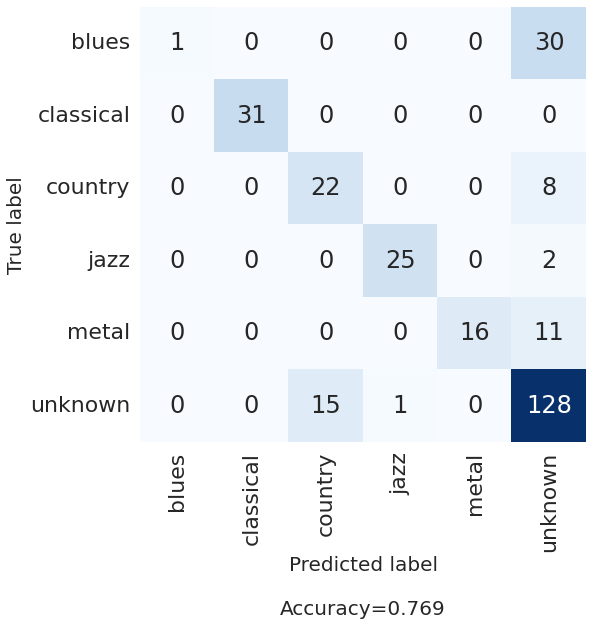}
    \caption{SoftMax with $\epsilon=0.4$.}
    \label{GTZAN_threshold_confu}
\end{subfigure}
~ 
\begin{subfigure}[hbt!]{0.46\columnwidth}
    \centering
    \includegraphics[width=\textwidth]{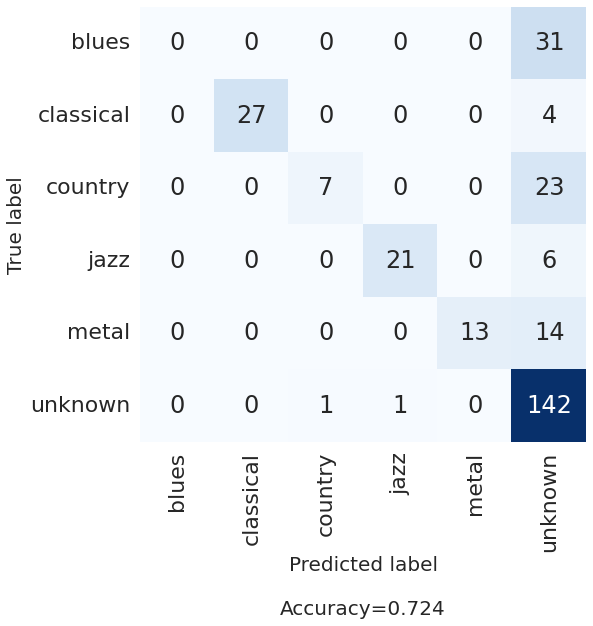}
    \caption{OpenMax with $\alpha=2$, $\eta= 20$, and $\epsilon=0.3$.}
    \label{FMA_small_conf}
\end{subfigure}
~ 
\caption{Confusion matrix for the GTZAN test split using the SoftMax with threshold and the OpenMax with treshold. }
\label{fig:gtzan_confuse}
\end{figure}

Comparing the performance of SoftMax and OpenMax with threshold we note that in plot (a) Figure \ref{fig:threshold}, SoftMax with threshold has the highest optimal accuracy, though OpenMax performs best in lower $\epsilon$ regimes. In (b), SoftMax with threshold outperforms OpenMax but they converge to a similar maximum accuracy. In both (c) and (d) OpenMax consistently obtains higher accuracy than SoftMax indicating that OpenMax likely performs best when the closed set classifier has been trained on a larger dataset. 
Inspecting the confusion matrices in Figure \ref{fig:gtzan_confuse} we also see that OpenMax is slightly less prone to misclassifying Country as unknown.

Regarding the impact of dataset size, Table \ref{tab:performance} shows that both $ACC_C$ and $ACC_{ot}$ accuracies are higher for the FMA large experiments than those on the FMA medium.
In addition, genre classes with too few samples are unlikely to provide rich enough data to the closed set genre classifier to learn meaningful embeddings that would also be effective in the open set task.
Specifically, the model struggles to effectively classify the genres with limited samples in training. For example, we have extremely low precision on Blues and Country and can see from Table \ref{tab:fma_large_sample} that these classes have the smallest training dataset sizes, which is likely impacting the model precision.


\begin{figure*}[hbt!]
\centering
  \begin{subfigure}{0.33\textwidth}
    \centering
    \includegraphics[width=.95\linewidth]{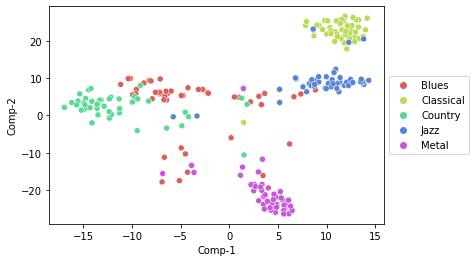}
    \caption{The t-SNE plot of KKCs.}
    \label{fig:1_tsne}
  \end{subfigure}%
  \begin{subfigure}{0.33\textwidth}
    \centering
    \includegraphics[width=.95\linewidth]{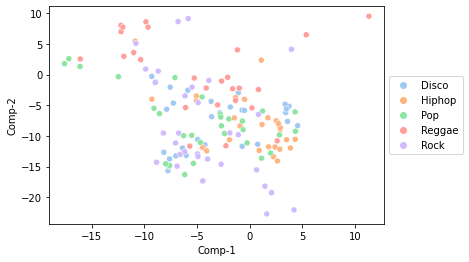}
    \caption{The t-SNE plot of UUCs.}
    \label{fig:2_tsne}
  \end{subfigure}
  \begin{subfigure}{0.33\textwidth}\quad
    \centering
    \includegraphics[width=.95\linewidth]{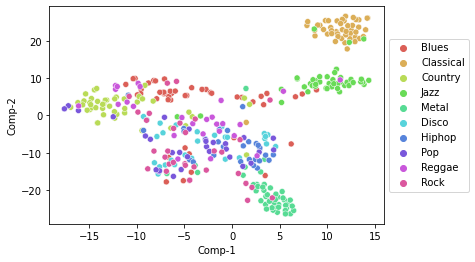}
    \caption{The t-SNE plot of both KKCs and UUCs.}
    \label{fig:3_tsne}
  \end{subfigure}
  \medskip
    \begin{subfigure}{0.33\textwidth}
    \centering
    \includegraphics[width=.95\linewidth]{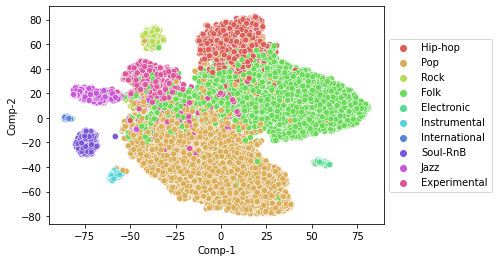}
    \caption{The t-SNE plot of KKCs.}
    \label{fig:4_tsne}
  \end{subfigure}%
  \begin{subfigure}{0.33\textwidth}
    \centering
    \includegraphics[width=.95\linewidth]{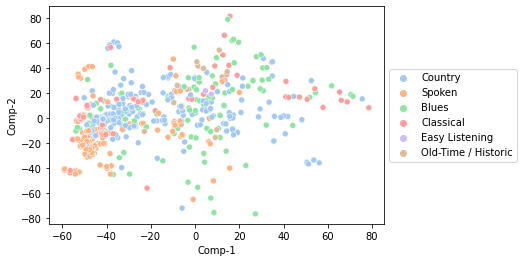}
    \caption{The t-SNE plot of UUCs.}
    \label{fig:5_tsne}
  \end{subfigure}
  \begin{subfigure}{0.33\textwidth}\quad
    \centering
    \includegraphics[width=.95\linewidth]{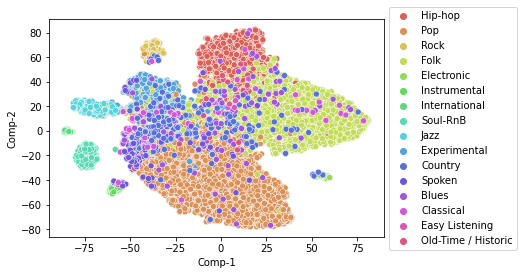}
    \caption{The t-SNE plot of both KKCs and UUCs.}
    \label{fig:6_tsne}
  \end{subfigure}
  \medskip
    \begin{subfigure}{0.33\textwidth}\quad
    \centering
    \includegraphics[width=.95\linewidth]{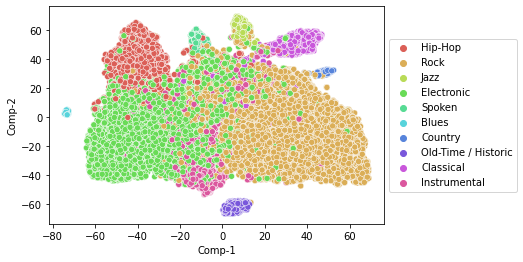}
    \caption{The t-SNE plot of KKCs.}
    \label{fig:7_tsne}
  \end{subfigure}
    \begin{subfigure}{0.33\textwidth}
    \centering
  \includegraphics[width=.95\linewidth]{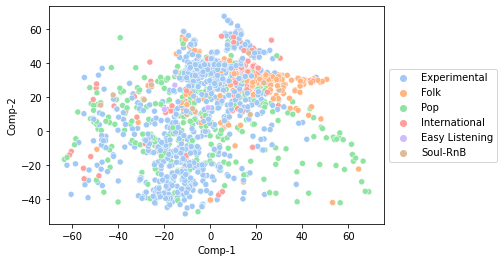}    \caption{The t-SNE plot of UUCs.}
    \label{fig:8_tsne}
  \end{subfigure}%
  \begin{subfigure}{0.33\textwidth}
    \centering
    \includegraphics[width=.95\linewidth]{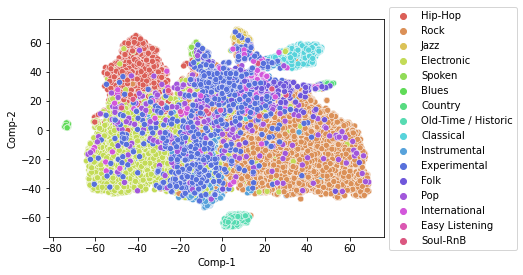}
    \caption{The t-SNE plot of both KKCs and UUCs.}
    \label{fig:9_tsne}
  \end{subfigure}
  \caption{The t-SNE plots. (a)-(c) are the t-SNE plots for Experiment 1. (d) - (f) are the t-SNE plots for Experiment 2. (g) - (i) are the t-SNE plots for Experiment 4. The t-SNE plots for Experiment 3 is not presented as it is similar to the plot for Experiment 4 due to the fact that they have the same KKCs and UUCs. }
\label{fig:tsne}
\end{figure*} 

Experiments 2 and 3 illustrate the impact that the choice of closed and open set classes has on the OSR model accuracy. Colloquially, less a clearly defined genres like Pop and Experimental were exchanged with Jazz and Classical in the open set. The hypothesis that the classes moved to the open set in Experiment 3 are more challenging to define based on audio features is consistent with the higher $ACC_C$ in experiment 3. Both $ACC_{st}$ and $ACC_{ot}$ are also higher, providing further evidence that a `best-case' closed set classifier will provide the optimal accuracy when including the open set in the task.

The t-SNE plots of the activation vectors of each song in Figure \ref{fig:tsne} also illustrate how much more densely concentrated around their centroid some genres like Classical are. In clear juxtaposition to this, Pop labeled samples are far less concentrated and overlap significantly with the spaces predominantly populated by other genres.

Looking more closely at specific genres, we have already noticed that most audio samples labeled as Blues are misclassified as unknown in the open set recognition task, as shown in Figure \ref{fig:gtzan_confuse}.
This is consistent with research found in \cite{sturm2013gtzan} that there are manually mislabeled audio samples between Blues and Country. In addition, there are mislabelled data points for the Metal genre label. Plot (a) in Figure \ref{fig:tsne} illustrates a number of the outliers for different classes, such as the Blues activation vectors near the Metal cluster. The combination of limited data points labelled Blues, outliers, and highly distributed samples could be resulting in probabilities of all KKCs, thus resulting in all Blues samples being labeled as unknown. 

Overall, the plots in Figure \ref{fig:tsne} indicate how distributed and less tightly defined the UUCs are than KKCs. In part this is to be expected because the activation vectors visualized are optimal representations learned from audio to separate the samples of each KKC. Pop and Hip-hop are significantly overlapping, and are also distributed over Classical, Jazz and Metal clusters. However, even some of the UUCs are more tightly distributed than others. For example, the Classical samples in plot (e) of Figure \ref{fig:tsne} are noticeably nearer to each other than those for Blues. A genre that can be succinctly defined in terms of the features in its audio is likely to have higher precision, even if belonging to the set of UUCs.


\begin{table}[hbt!]
\renewcommand{\arraystretch}{1.2}
  \centering
  \begin{adjustbox}{width=0.43\textwidth}
  \begin{tabular}{|c|c|c|c|}
  \hline
    Genre & Training sample & Precision & Recall\\
\hline 
    Hiphop & 2939 & 0.407 & 0.667   \\ 
\hline 
    Rock &11386 & 0.581 & 0.914 \\
\hline 
    Electronic & 7649 & 0.254 & 0.860\\ 
\hline 
    Instrumental &1578 & 0.031 & 0.091\\
\hline 
    Jazz & 462 & 0.387& 0.176 \\
\hline 
    Blues & 88 & 0.000 & 0.000\\
\hline 
    Spoken &276& 0.597 & 0.254\\
\hline 
    Country  &158& 0.000 & 0.000\\
\hline 
    Classical &1069& 0.784 &0.453\\
\hline 
    Old-time Historic &440& 0.556 & 0.789\\
\hline
    Unknown & None&  0.822 &  0.627\\
\hline
  \end{tabular}
  \end{adjustbox}
  \caption{
  This table contains the training data size, precision and recall for each genre in the FMA large experiment. It also includes the precision and recall for the UUCs that were not present in training.
  }
  \label{tab:fma_large_sample}
\end{table}

\begin{figure*}[h!]
\centering
\begin{subfigure}[h!]{0.99\columnwidth}
    \centering
    \includegraphics[width=0.9\textwidth]{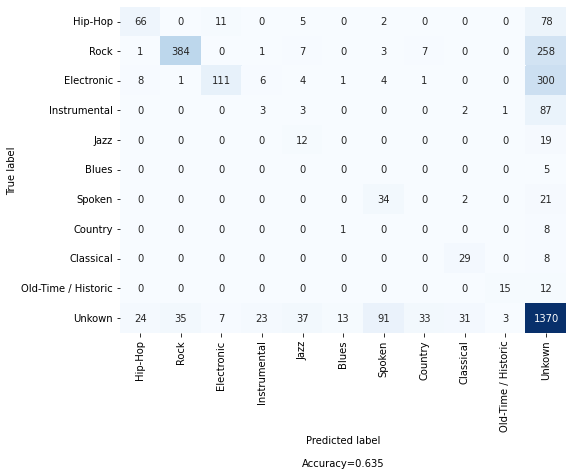}
    \caption{Confusion matrix using OpenMax with threshold.}
    \label{FMA_large_confusion}
\end{subfigure}
~ 
\begin{subfigure}[h!]{0.9\columnwidth}
    \centering
    \includegraphics[width=0.95\textwidth]{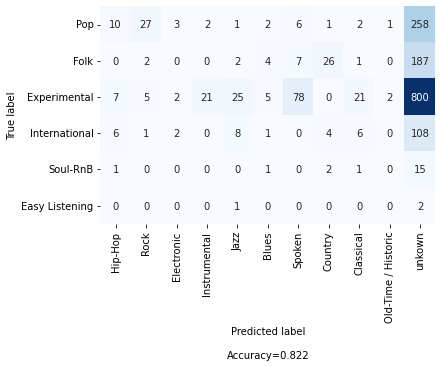}
    \caption{Error matrix using the OpenMax with threshold. }
    \label{FMA_large_open_matrix}
\end{subfigure}
~ 
\caption{Confusion matrix and error matrix for the experiment using the FMA large dataset. (a) is the confusion matrix showing how each genre category in the test split is predicted. (b) is the error matrix illustrates on how the unknown genre categories predicted incorrectly into close set genre categories or correctly as the unknown genre category. }
\label{fig:fma_large}
\end{figure*} 



\section{Conclusion \& Future Work}
\label{sec:con}

This paper establishes a baseline capacity and viable approach to novel genre detection by applying open set recognition methods to embeddings learned from `best-case' closed set genre classifiers. We propose a system for open set recognition on music genres, which is both capable of classifying audio into known genres and also detecting previously unseen genres. We conduct a full experimental analysis of its performance on experiments designed from open source benchmark datasets and find that the open set  accuracy can be affected by factors including the rejection threshold, the number of training samples, the choice ofopen set recognition model, the selection of KKCs and UUCs, and the data label quality, influenced by the qualitative definition of genres. 


As \cite{sturm2013gtzan} suggested, there are issues with labeled genre tags, even those provided by experts in the GTZAN dataset. There are similar caveats and complications in the artist provided labels in the FMA dataset \cite{zhang2019songnet}. A possible research direction is to better define music genres based on audio features and design machine learning systems to automatically correct mislabeled genres, leveraging unsupervised approaches to learning groups of similar music. We could also employ few-shot learning to identify music genres for which we have limited training data. Furthermore, genres emerge over time, and a practical system that detects the emergence of novel genres from audio could be trained on samples of genres that are segmented into KKCs and UUCs based on when they were released, allowing the trained model to provide a realistic measure of our ability to identify if music is of a new genre.


\clearpage
\bibliography{ISMIRtemplate.bib}

\end{document}